# Two orders of magnitude improvement in detection limit of lateral flow assays using isotachophoresis

*Babak Y. Moghadam†, Kelly T. Connelly‡, Jonathan D. Posner†\**

†University of Washington, Mechanical Engineering Department, Seattle, WA 98195, USA

‡University of California, Los Angeles, Mechanical and Aerospace Engineering Department, Los Angeles, CA 90095, USA

**Abstract**

Lateral flow (LF) immunoassays are one of the most prevalent point-of-care (POC) diagnostics due to their simplicity, low cost, and robust operation. A common criticism of LF tests is that they have poor detection limits compared to analytical techniques, like ELISA, which confines their application as a diagnostic tool. The low detection limit of LF assays and associated long equilibration times is due to kinetically limited surface reactions that result from low target concentrations. Here we use isotachophoresis (ITP), a powerful electrokinetic preconcentration and separation technique, to focus target analytes into a thin band and transport them to the LF capture line resulting is a dramatic increase in the surface reaction rate and equilibrium binding. We show that ITP is able to improve limit of detection (LOD) of LF assays by 400-fold for 90 second assay time and by 160-fold for a longer 5 minutes time scale. ITP-enhanced LF (ITP-LF) also shows up to 30% target extraction from 100 μL of the sample, while conventional LF captures less than 1% of the target. ITP improves LF assay LOD to the level of some lab based immunoassays, such as ELISA, and may provide sufficient analytical sensitivity for application to a broader range of analytes and diseases that require higher sensitivity and lower detection limits.

## 1. Introduction

The emphasis of medical care is shifting toward prevention and early detection of disease and management of multiple chronic conditions resulting in higher prevalence of point-of-care (POC) testing.[1] Introduced in 1987, lateral flow (LF) immunoassay are one of the most commercial available rapid POC tests.[2–5] POC lateral flow-based tests have enjoyed commercial success because of their simplicity, low cost, and robust operation. LF tests use a paper substrate, e.g. porous membranes, to indicate presence of target analytes. In a conventional LF assay, a target analyte binds to a labeled detection protein, e.g. antibody labeled with colloidal gold, and travels across the membrane by capillary action, without user's manipulation. At the test zone, the target binds with a specific capture molecule immobilized on the surface which accumulates the captured target and generates a visually detectable, i.e. colorimetric, signal.

LF tests have been indicated as a suitable medical diagnostic tool for remote or impoverished settings and are also used as an appropriate technology for a wide variety of field applications.[6] However, conventional LF assays have relatively poor detection limits ($\sim 10^{-11}$ M)[7,8] compared to analytical methods like ELISA and SPR ($\sim 10^{-12} - 10^{-18}$ M).[9] In many diagnostic situations, especially at the early stages of a disease, protein markers, such as pathogen antigens or host antibodies, are present at very low concentrations. The serum concentrations of the majority of proteins important in cancer,[10] neurological disorders[11,12] and the early stages of infection disease[13] are reported to range from $10^{-12}$ to $10^{-16}$ M. LF tests typically fail to provide a reliable and consistent detection at these target concentrations while lab-based analytical techniques, like ELISA, can routinely detect at these levels and have been shown to operate at concentrations as low as $10^{-19}$ M.[14] Moreover, LF tests are often single step assays and are specific to antibody-antigen interaction which limits their application for integrated sample preparation (e.g. such is required for nucleic acid purification) and a narrow group of samples.

The vast majority of work toward improving detection limits of LF assays has been focused on optimizing affinity of the antibodies used in the test,[15,16] and introducing new labeling techniques.[17–19] In optimal physiochemical conditions (pH, ionic strength, and temperature) the affinity, $K_d$, of a specific antibody-antigen pair has a value that is constant and to a large extent influences the analytical sensitivity of the test.[6] One useful metric in quantifying the detection limit of an immunoassay is the sample target concentration $C_T$ normalized by the affinity constant $K_d = k_{off}/k_{on}$, where $k_{off}$ and $k_{on}$ are respectively the off- and on- rate constants of the binding reaction. The lowest normalized detection concentration $C_T/K_d$ reported as ~$10^{-2}$ for standard LF tests[8] and for ELISA the value drops to the range of $10^{-5}$ to $10^{-10}$. A low value of $C_T/K_d$ suggests a lower concentration detection limit. Much effort has been focused on decreasing $K_d$ values of secondary antibodies, in the range of $10^{-6}$ to $10^{-10}$,[20] to reduce the concentration detection limits of immunoassay, including LF tests. Alternatively, the detection limit of a LF test can be improved by a factor of nearly 10-fold by using different labels and labeling techniques.[17,18] However, to improve labeling, usually a high concentration of the labeling reagent is necessary which can increase final cost of the assay. Moreover, conjugating the antibodies with larger labeling molecules can lead to a change or even a complete loss of affinity due to the effect of steric restrictions.[6] Recently, there has been a push to enhance the signal in LF assays by developing paper-based microfluidic devices that incorporate sophisticated functions and device designs using capillary action to drive the flow.[21–23] In spite their success in integrating multiple functions and assays on a single paper device, detection limit improvement reported in these studies is limited to about 3-fold. Furthermore, these methods add more complexity and cost to the final product which may limit their application in POC settings.

Improving reagent affinity, labeling techniques, and complex device features have not alone addressed the challenge of detecting low target concentrations ($10^{-12}$ to $10^{-16}$ M) observed in diseases of interest. At these target concentrations the $K_d$ for a LF assay would need to be reduced to a value near that of the strongest affinity of $10^{-15}$ reported for streptavidin-biotin interaction[24,25] to reach a $C_T/K_d$ of $10^{-2}$. Surface binding of the antibody-antigen in LF assays can be regarded as a second-order reaction where the binding rate depends on the reaction rate constants and concentration of the reactants. At low target concentrations the forward reaction rate is decreased, i.e. reaction is kinetically limited, and as a result it takes impractically long reaction times for the assay to equilibrate and generate a detectable signal. Increasing the surface concentration of the immobilized capturing reagents can improve the forward reaction rate, but is ultimately limited by the protein binding capacity of the membrane. Therefore, to improve the detection limits and slow binding rates in an immunoassay, preconcentration of the target becomes a necessary step.[26–28] A number of strategies are currently available to improve detection limits of the surface reaction in biosensors using preconcentration, including solid-phase extraction,[29,30] electrokinetic techniques,[31,32] as well as chromatographic and membrane preconcentration.[33–35] Electrokinetic techniques address the target preconcentration with minimal requirements and pretreatment of the sample. Isotachophoresis (ITP) preconcentration has been recently used to accelerate and improve detection limits of surface DNA hybridization assays. In these studies surface-based reaction in microchannels was accelerated by preconcentrating the target analytes and transporting them to the test zone which considerably improved the assay detection limits.[36,37]

ITP is a powerful electrokinetic technique used widely in capillaries and microchannels for the preconcentration, separation, and purification of ionic analytes.[38] Under an applied electric field during ITP, target analytes transport electrophoretically and accumulate at the boundary of fast moving leading (LE) and slow moving trailing electrolyte (TE) ions, resulting in substantial increase in analyte concentration in a narrow plug, on the order of 100 μm wide. ITP in microchannels has demonstrated up to a million-fold increase in target concentration during the assay time of less than 30 minutes.[39] We previously showed that ITP can increase the width-averaged concentration of target analytes in nitrocellulose membrane strip by up to 900-fold in less than 5 minutes.[40,41] We also demonstrated that ITP on nitrocellulose can extract more than 80% of the target from initial sample, is compatible with relatively large sample volumes (>100 μL) and can be powered by a portable power supply showing its potential to be used in POC devices. It has been shown that ITP can enhance the surface DNA microarray hybridization reaction between a suspended target and an immobilized probe in conventional microfluidic devices. Han et.al demonstrated the acceleration and sensitivity improvement of DNA microarray hybridization by preconcentrating the target molecules and transport them over the immobilized probe site using ITP.[37] Their method leverages preconcentration of ITP to overcome the slow reaction kinetics of surface hybridization. Their analytical predictions along with experimental observations show 8-fold improvement in the detection limit at about 30-fold shorter assay time.

In this paper, we use ITP preconcentration of antibodies on nitrocellulose membranes to improve limit of detection (LOD) of LF assays. ITP increases concentration of the target molecules and transports them in a thin band to the test zone which results in an increase in kinetic reaction rate of the target and the probe and more than two orders of magnitude improvement in the LOD. We use the analytical model, developed by Han et al.,[37] for the ITP-enhanced second order surface reaction and apply it to our system. We perform quantitative fluorescence ITP-enhanced lateral flow (ITP-LF) assay on nitrocellulose membranes and compare its performance to the conventional LF assay. We also use antibodies labeled with colloidal gold as our target to investigate compatibility of ITP-LF with colorimetric detection, which is used in conventional LF tests for visual interpretation of the test results. Finally, we demonstrate an improvement of detection limit in mouse serum showing that ITP can extract and preconcentrate targets present in biological media.

## 2. Theoretical model

The basic principle of any immunochemical technique, including LF immunoassays, is that a specific target, e.g. an antigen, binds with its specific probe, e.g. a secondary antibody, to give an exclusive target-probe complex. The specific binding of antibodies is dependent on hydrogen bonds, hydrophobic interactions, electrostatic forces, and van der Waals forces, therefore it is a reversible process. For a single probe (P) binding site that binds to a single target (T) and forming a complex (C), the surface reaction can be modeled as a second-order reaction with on- and off- rate constants of respectively $k_{on}$ and $k_{off}$, and the dissociation constant of $K_D = k_{off} / k_{on}$.[42]

By assuming perfect mixing conditions (where the reaction is kinetically limited), and negligible consumption of the target molecules at the test zone, one can develop a relation for this reaction. In this regime, target concentration above the immobilized probe remains unchanged from its initial value, $C_{T0}$.[37,43,44]

$$h_{LF} = \frac{\tilde{C}_C}{\tilde{C}_{P0}} = \frac{C_0^*}{C_0^* + 1}\left(1 - \exp\left(-\left(C_0^* + 1\right)k_{off}t\right)\right) \quad (1)$$

Here $h_{LF}$ is the fraction of bound probe, $C_C$ and $C_{P0}$ are molar surface concentrations of the bound probe and of the probe at $t = 0$. $C_0^*$ is defined as normalized initial target concentration, $C_{T0}/K_D$ where $C_{T0}$ is the volumetric molar concentration of the suspended target in the immediate vicinity on the immobilized probe in the test zone.

ITP-LF assay can be similarly modeled as a kinetically limited reaction at an elevated target concentration resulted from the ITP preconcentration. This means that the concentration of the target in the ITP zone stays constant above the probe area at all times. We can assume that the ITP-LF is kinetically limited because the typical ITP residence time, i.e. time during which probe area is exposed to the ITP zone, $t_{ITP}$ = 284 sec here, is much longer than the diffusion time scale. Diffusion time scale for the target to diffuse to the probe surface in nitrocellulose membrane can be estimated as $t_{diff}$ = 0.32 sec using $t_{diff} = d^2/D$ where $d$ is the pore diameter of the membrane (8 μm) and $D$ is the diffusivity of IgG secondary antibody in nitrocellulose membrane ($2\times10^{-10}$ m$^2$/s).[20] ITP is also known to cause secondary flows which help mitigate diffusion-limited regimes.[45]

Han et al. showed that by modeling the Gaussian concentration profile of the ITP-focused sample as a top-hat pulse with characteristic ITP zone width of $\delta_{ITP} = \pm 2\sigma$ and target concentration of $pC_{T0}$, where $p$ is the effective preconcentration ratio and $\sigma$ is the standard deviation of the sample Gaussian distribution, an approximate analytical model for ITP enhanced surface binding can be derived as,[37]

$$h_{ITP-LF} = \frac{pC_0^*}{pC_0^*+1}\left(1-\exp\left(-\left(pC_0^*+1\right)k_{off}t_{ITP}\right)\right), \tag{2}$$

where $h_{ITP-LF}$ is the fraction of bound probe in ITP-LF. They derived this equation by substituting $pC_{T0}$ for $C_{T0}$ and $t_{ITP}$ for $t$ in equation (1). The ITP plug migrates at a constant velocity of $V_{ITP}$, therefore each point at the test zone is exposed to the highly focused sample for $t_{ITP} = \delta_{ITP}/V_{ITP}$. $p$ is calculated such that the area under the top-hat profile, $pC_{T0}\delta_{ITP}$, is the same as the area under the Gaussian distribution of the target in the ITP plug, $\sqrt{2\pi}A\delta_{ITP}/4$, where $A$ is the maximum value. We use equations (1-2) to compare the signal strength, LOD and assay time of the ITP-LF and conventional LF assay. This theory gives us an insight to better understand the design and optimization of ITP-enhanced LF assays.

## 3. Material and methods

### 3.1. Reagents and the electrolytes system

We performed a series of quantitative and qualitative ITP experiments on nitrocellulose membranes demonstrating the improvement of LF assay LOD using ITP. We used IgG secondary antibodies as the target and capture reagents since they are extensively used in LF assays as the capturing and labeling reagents. Moreover, these antibodies have high binding affinity for nitrocellulose membrane therefore higher surface concentration of the immobilized probe can be achieved which improves the capturing. For target analyte we used goat anti-rabbit IgG labeled with Alexa Fluor 488 fluorescence dye (CAS# 2628-2-8, Molecular Probes, Eugene, OR), for quantitative detection, and goat anti-mouse IgG labeled with 40 nm colloidal gold (Arista Biologicals, Allentown, PA) for colorimetric detection. We used bovine anti-goat IgG (Jackson ImmunoResearch Laboratories, West Grove, PA) as the capturing reagent, i.e. probe, since it recognizes and binds with the goat IgG and has minimal cross reactivity for bovine albumin serum (BSA) that we used to block the membrane.

To design the ITP electrolyte system we first measured the effective electrophoretic mobility of the target IgG in a wide range of pH (data shown in SI). IgG antibodies are large molecules of about 150 kDa thus they show low electrophoretic mobilities especially around the neutral pH, due to their isoelectric point. Based on the measured effective mobility of IgG as 6.3 nm$^2$/Vs (at pH 7.4) we designed the TE composition as 10 mM Glycine, with the effective mobility of 0.4 nm$^2$/Vs, buffered with Bis-Tris to pH 7.4 ($\sigma$ = 3 µS/cm). At this pH the binding reaction occurs at a biologically relevant condition and the difference between the mobilities of target and TE ions is maximized, which results in more potent sample focusing. We used semi-finite sample injection by mixing the TE with target analyte at different concentrations. The LE contained 40 mM HCl buffered with Tris to pH 8.1 to maintain high buffering capacity of the Tris ($\sigma$ = 4.2 mS/cm). The conductivity, $\sigma$, ratio of the LE and TE was maximized to 1400 to achieve the highest preconcentration while minimizing the Joule heating. We added 0.5% PVP to the LE to suppress electroosmotic flow. For a detailed discussion on the choice of electrolyte system in ITP, refer to the works of Bagha et al. and Moghadam et al.[40,46] All the reagents were purchased from Sigma Aldrich (St. Louis, MO) unless mentioned otherwise. All aqueous samples were prepared using water ultrapurified with a Milli-Q Advantage A10 system (Millipore Corp., Billerica, MA).

### 3.2. Paper device preparation

We used backed nitrocellulose membrane (HF-125, Millipore Corp., Billerica, MA) cut with a CO$_2$ laser ablator (Universal Laser Systems, Scottsdale, AZ) to fabricate the immunoassay devices used in both LF and ITP-LF experiments. Paper device dimensions were chosen as 40 mm × 3.5 mm to maximize the target accumulation in the ITP zone while reducing the Joule heating caused by high electrical resistance of the device.[40] We immobilized the capture reagent on the membrane by using a custom-made protein spotting system.[47] To maintain physiological condition, capture IgG was suspended in 0.25 M NaCl and 0.01 M sodium phosphate, buffered to pH 7.0. This pH is close to the isoelectric point of the IgG, which helps to destabilize the probes in aqueous form therefore they bind more effectively with the membrane surface. The test line was located one-third of the device length away from the LE side where higher sample preconcentration ratio is reached. We spotted the test line with the width of approximately 800 µm and measured probe surface density of $C_{P0} = 2 \times 10^{-9}$ moles per internal surface area of the membrane (0.5 fmole of probe per test line).[47] Thickness of the test line was chosen to be in the same order of the ITP plug width, $\delta_{ITP}$, to increase the surface interaction and $C_{P0}$ was chosen close to the IgG binding capacity of the nitrocellulose reported by Fridley *et al.*'s measurements.[20] After spotting, the membranes were allowed to dry in room temperature for 5 minutes then further dried at 37º C for 1 hour. To minimize the non-specific interaction of the target antibody with the membrane, we used 1% BSA as the non-reactive blocking agent.[48]

### 3.3. LF and ITP-LF protocol

Figure 1A shows our ITP experimental setup where the paper device is placed in an acrylic holder with the test zone closer to the LE reservoir. Before each ITP experiment the membrane was hydrated with the LE and placed on the holder with the left and right reservoirs filled respectively with 100 µL of the TE, mixed with the sample, and LE. To apply electric field across the membrane we placed the positive and negative platinum wire electrodes in the LE and TE reservoirs, respectively. We initiated the ITP-LF assay by applying 500 µA constant current to the TE reservoir and grounding the LE using a high voltage power supply (HSV488 6000D, LabSmith Inc., Livermore, VT). Higher currents result in a higher preconcentration ratio and a shorter assay time,[40,49] but lower the signal strength at the test zone because the ITP zone residence time over the test line is decreased. Therefore, as the ITP zone arrives at the test zone we reduced the current to 50 µA which reduces the ITP velocity thus increases the reaction time. Plug velocity, and therefore the ITP zone residence time, is inversely proportional to the applied current.[49] For example, in our system, $t_{ITP}$ = 9 sec at 500 µA, and $t_{ITP}$ = 284 sec at 50 µA. After the ITP plug crosses the test zone we increased the current back to 500 µA until the plug reaches the LE reservoir. Progression of the ITP plug downstream subsequently acts as an *electrokinetic wash* step where unbound targets are removed from the probe spots and collected at the migrating ITP plug so no further washing step is required.[36,37] For the conditions mentioned above, each ITP experiment takes about 7 minutes to complete.

For conventional LF assays we used a holder with one reservoir, as shown in Figure 1B. We placed one end of the membrane in the reservoir and the other end on a cellulose absorbent pad (HF-125, Millipore Corp., Billerica, MA) to wick the sample during the assay. To initiate the LF assay we filled the reservoir with the target solution mixed with TE which triggers the capillary action and movement of sample across the membrane. The absorbent pad continuously wicks the sample allowing constant flow of the target analytes during the experiment therefore target molecules interact with the immobilized probe throughout the duration of the test. We let the interaction progress for a certain amount of time, then replaced the sample solution with DI water to rinse the membrane and remove the unbound target.

### 3.4. Detection and quantification

We performed quantitative fluorescence experiments using a Nikon AZ100 upright microscope equipped with a 0.5x objective (Plan, NA 0.5, Nikon Corporation, Tokyo, Japan), an epifluorescence filter cube (488 nm excitation, 518 nm emission, Omega Optics, Brattleboro, VT), and a 16-bit, cooled CCD camera (Cascade 512B, Photometrics, Tucson, AZ) which capture images at the exposure time of 1 s. Images taken from the test zone after each experiment were processed using an in-house MATAB code to quantify the amount of target captured in the test area, $m_T$, effective stacking ratio, $p$, and ITP residence time, $t_{ITP}$. We calculated both the mean and standard deviation of the background, determined from a cropped image of the device excluding the test line. The mean-subtracted intensity of each pixel in the test area was then counted as the signal if it was greater than 3 standard deviations of the background, i.e. signal to noise ratio (SNR) of 3. To convert the signal intensity values to the target concentration, we generated a calibration curve by performing three titration experiments. In each titration experiment we measured fluorescence intensity of the membrane surface fully wetted by a known high concentration of the target solution and generated a calibration curve from a linear regression (example is shown in the SI). The calibration experiments were done after performing each set of experiments, on multiple days, and we found the results to be very consistent and repeatable.

## 4. Results and discussion

### 4.1. Demonstration of ITP enhanced LF

After applying the electric field to the system, a highly focused plug of target molecules electromigrates to the test zone where they react with capture molecules immobilized on the surface. ITP increases the target concentration and interaction kinetics of the target-probe which results in generation of a specific fluorescence signal at the test zone. In Figure 1C, we present five snapshots of a single ITP-LF experiment showing successful capturing of the target using ITP (an example video of this experiment is provided in the SI). In snapshots a-c, an ITP zone containing highly concentrated target forms and migrates across the membrane. As the ITP zone travels downstream it sharpens and its fluorescence intensity increases demonstrating that the target concentration is increasing. As the plug approaches the test zone, we decrease the applied current and the slow moving plug sweeps over the test zone and binds with the immobilized probes. After the ITP zone passed the test zone (snapshots d-e), we observe a fluorescence signal at the test zone indicating that the target molecules have been captured by the immobilized probes. Any unreacted target molecules electromigrate towards the LE reservoir which effectively serves as a membrane wash step.

### 4.2. Experimental determination of the kinetic and focusing parameters

Kinetic parameters, $K_D$ and $k_{off}$, as well as the ITP focusing parameters, $p$ and $t_{ITP}$, need to be determined in order to compare the experimental results with the model predictions described in equations 1 and 2. To obtain the kinetic parameters we performed a conventional LF assay experiment using 25 mg/l initial target concentration and monitored the signal over time. Figure S-3 of the SI shows the captured amount of target in LF assay as a function of time. We fit Equation 2 to these data using the least square method using $K_D$ and $k_{off}$ as the fitting parameters. The fitting parameters were determined as, $k_{off} = 1.75 \times 10^{-3}$ s$^{-1}$ and $K_D = 1.42 \times 10^{-6}$ M which are consistent with the range of values reported for the interaction of secondary antibodies.[50,51]

We obtained the ITP focusing parameters, $p$ and $t_{ITP}$, by conducting independent ITP experiments. We performed ITP experiments and measured the ITP zone velocity, $V_{ITP}$, width of the ITP plug, $\delta_{ITP}$, and the effective preconcentration ratio, $p$. For each ITP experiment we obtained images of the ITP plug prior to reaching the test zone, then converted the intensity values to target concentration using our calibration curve. We fit a Gaussian distribution in the form of $A\exp(-(x-\mu)^2/2\sigma^2)$ to the resulting concentration profile.[49] The resulting fitting parameters are the maximum plug concentration $A$, mean $\mu$, and the plug width $\delta_{ITP} = \pm 2\sigma$ (two times the plug standard deviation). We experimentally estimated the effective ITP preconcentration ratio of $p = 92 \pm 12$ for $I = 500$ µA using a method described earlier. The ITP velocity was measured by dividing the average displacement distance of the Gaussian fit obtained from twenty pairs of images taken with 1 second image-to-image delays. By dividing the value of $V_{ITP}$ to $\delta_{ITP}$ we obtained the characteristic ITP reaction time as $t_{ITP} = 9.4 \pm 0.74$ s for $I = 500$ µA, and $t_{ITP} = 284 \pm 47$ s for $I = 50$ µA. The plus/minus values represent 95% confidence interval obtained from 3 measurements. We used the later value for our model since the current is set to 50 µA when the ITP zone is crossing the test zone in ITP-LF.

**4.3. Capture Enhancement by ITP**

Figure 2 shows experimental data for the fraction of bound probe $h$ as a function of target concentration for ITP-LF and LF. Assuming one-to-one binding of target and probe, the fraction of bound probe is defined as the surface concentration of the bound probe, $C_C$, normalized by the initial surface concentration of the probe, $C_{P0}$. We estimated the fraction of bound probe by normalizing the amount of bound target in the test zone with its maximum value. Each ITP-LF takes about 7 minutes to complete with the ITP zone residence time of 284 s. LF data were collected at three different assay times of 90 seconds, 5 minutes and 1 hour. $h$ increases with initial target concentration for both LF and ITP-LF. The amount of captured sample in ITP-LF increases with increasing target concentration, however at concentrations higher than 25 mg/l we did no observe a change in the amount of captured target which suggests that the probe surface was saturated. At each concentration the fraction of bound probe using ITP is higher than the conventional LF since the ITP preconcentration increases the reaction rate. For example, at 10 mg/l the fraction of bound probe is $6.7 \times 10^{-1}$ for ITP-LF and is $8.53 \times 10^{-3}$ for LF after 5 min. This shows improved capture by ITP preconcentration by about 80 fold. $h_{LF}$ increases with time and reaches near equilibrium conditions after one hour, which can be seen by taking the derivative of Equation 1 in respect to $t$.

Along with the experimental data in Figure 2 we present analytical models (equations 1-2) for ITP-LF and conventional LF. To obtain these theoretical curves, we used the experimentally determined kinetic and focusing parameters as $k_{off} = 1.75 \times 10^{-3}$ s$^{-1}$, $K_D = 1.42$ μM, $p = 92$ and $t_{ITP} = 284$ s. For LF, we observe good agreement of predicted trends and our experimental data. For ITP-LF, we notice good agreement at higher concentrations but lower than predicted binding at 0.05 mg/l. We hypothesize that this over prediction of $h$ by the model at low concentration may be due to deviation from the kinetically limited assumption as a result of very low number of moles of target present above the reaction zone. The number of moles of target in the ITP zone here is approximately 8 fmoles. At this condition, roughly one-third of the target molecules have been consumed by the surface reaction, therefore the assumption of constant target concentration may not be valid.

LOD in our system is defined as the lowest target concentration at which the maximum background-subtracted intensity in the test zone is 3 standard deviations away from the background, i.e. SNR = 3.[52] In practice, for example, the normalized standard deviation for ITP-

LF is ~ $8\times10^{-4}$ and the fraction of bound target, $h$, corresponding to LOD is ~ $5.4\times10^{-3}$ for ITP-LF and ~ $10^{-2}$ for LF. The bound target concentration of ITP-LF at LOD is marginally lower because the electrokinetic "wash" step during ITP effectively removes the unbound target from the surface and improves the SNR. Table 1 lists the LOD extracted from Figure 2 for ITP-LF and LF at three assay times tested. This table shows that the LOD is improved by a factor of 60, 160, and 400 for 1 hour, 5 minutes, and 90 seconds LF assay times, respectively. The normalized initial target concentration, $C_{T0}/K_D$, at limit of detection is $3.3\times10^{-4}$ for ITP-LF which is two orders of magnitude below the lowest value reported for conventional LF assays.[8]

In addition to improving the detection limits, ITP increases the capturing efficiency of LF assays. In Table 2 we present the capturing ratio of ITP-LF, $(N_C/N_{T0})_{ITP-LF}$, as a function of initial target concentration, $C_{T0}$, and compare it to the capturing ratio of LF assay, $(N_C/N_{T0})_{LF}$. Capturing ratio is calculated by calculating moles of target captured in the test area, $N_C$, and normalizing it by the initial moles of sample in the sample reservoir, $N_{T0}$. Capture ratio for LF increases by increasing the initial target concentration and reaches 0.7 % at $C_{T0} = 25$ mg/l. Similar trend is observed for ITP-LF with the highest capture ratio of 30 % observed at $C_{T0} = 0.5$ mg/l. An apparent decrease in the capturing ratio of ITP-LF is observed at higher concentrations because the test line is saturated due to having excess amount of target molecules in the sample solution, i.e. $N_{T0}$ is larger than the initial moles of probe, $N_{P0}$. Unbound target molecules during ITP=LF travel with the ITP plug toward the LE reservoir as shown in Figure 1C-d. This suggests that we may be able to increase the capturing efficiency of ITP-LF by using even smaller volumes of the sample. LF assay shows capturing ratios smaller than 1% and we are not able to detect the target at concentrations lower than 8 mg/l which is below the LOD of LF assay. At each concentration, capturing ratio of the conventional ITP-LF assay is one order of magnitude larger than the capturing ratio of LF. This data suggests that integration of ITP with LF may offer the capability in using large sample volumes with very dilute concentration of target.

**4.4. Colorimetric detection in ITP-LF**

The capability of LF assays to generate an unambiguous and visually read result, i.e. colored band, is critical feature of their success as a POC diagnostic. This is achieved by using detection antibodies labeled with colloidal gold or other colorimetric signal generators such as colored latex particles,[53] and colloidal carbon nanoparticles.[54] In order to show the compatibility of our ITP-LF assay with colorimetric detection, we used IgG labeled with 40 nm colloidal gold (IgG-Au) as our target. We measured effective electrophoretic mobility of IgG-Au at pH 7.4 as 8.9 nm$^2$/Vs. Since this value is larger than the effective mobility of TE ions we used in our quantitative analysis, we kept the same TE composition to perform colorimetric ITP-LF experiments. Generation of a pink-colored band at the test zone indicates successful capturing of the target with the color intensity directly proportional to the amount of sample captured. ITP is able to focus IgG molecules labeled with gold nanoparticles and transport them across the membrane toward the test zone (a video of colorimetric ITP-LF is provided in the SI). Figure 3 shows snapshots of the paper device used for colorimetric detection of the target using both ITP-LF and conventional LF for the initial target concentration ranging from 0.1 – 15 mg/l. The optical density of the test line increases with initial target concentration as expected. Qualitatively, a visually detectable signal was observed using ITP-LF at concentrations as low as 0.1 mg/l while no signal was detected using LF at concentrations lower than 10 mg/l and 10 minutes assay time. Qualitatively, these data show more than 100-fold improvement in detection limit of LF assay using ITP which is consistent with our quantitative fluorescence measurements.

### 4.5. Insights into the design of an ITP-LF assay

We use our analytical model as a guide for the design of an ITP-enhanced LF assays. In Figure 4 we present the LOD ratio ($LOD_{LF}/LOD_{ITP-LF}$) versus the assay time for different effective preconcentration ratios and off-rate constants. Each value on the y-axis shows the improvement in LOD when using ITP at a given LF assay time. ITP preconcentration results in a considerable improvement in LOD of the LF assay due to the elevated forward reaction rate as a result of higher target concentration in the test zone. LOD ratio decreases at longer times since the binding of target and probe in kinetically limited LF assay increases with time. ITP improves the LOD of LF assay even at longer times despite the fact the LF reaches equilibrium. For example, at $p = 100$, ITP improves the LOD of LF assay by a factor of 30 at 1 hour assay times and longer. Figure 4 shows that for shorter assay time much higher LOD ratios can be obtained which means, in addition to the LOD improvement, ITP can speed up the LF reaction. In the SI we provide a discussion on increasing the reaction rate using ITP as was described in the work of Han et.al for DNA hybridization in microchannels.[37] Closed symbols in Figure 4 represent our experimental conditions where $p = 92$, $t_{ITP} = 284$ s and $k_{off} = 1.75 \times 10^{-3}$ s$^{-1}$, showing good agreement with the model predictions. We hypothesize that the under prediction by the model at 1 hour assay time is due to the fact that the LF may have not reached equilibrium. We believe that at very low target concentrations, LF is limited by the diffusion of the target molecules to the test zone and thus requires longer times, than predicted by the model, to reach equilibrium. Figure 4 also shows that at higher preconcentration ratios, e.g. $p = 500$, higher LOD ratios can be achieved, as expected.

We also examined effect of kinetic parameters and observed that the $k_{off}$ has marginal improvement effect on the LOD. For example, at 30 min assay time, increasing the $k_{off}$ by one order of magnitude increases the LOD ratio by less than 2-fold. When the $k_{off}$ is increased the $k_{on}$ value also increases, at constant $K_D$, therefore the forward reaction rate is increased. We observed that LOD ratio does not depend on the $K_D$ value because ITP improves the binding by increasing the forward reaction rate. The equilibrium dissociation constant is a representation of the affinity of two compounds and does not depend on time. For conventional LF assay, the fraction of the bound probe at longer times is directly proportional to the initial target concentration, $h_{LF} \approx C_{T0}/K_D$ for conditions satisfying $C_0^* \ll 1$. Similarly, in ITP-LF, a Taylor series expansion for the limiting case of $(pC_0^* + 1)k_{off}t_{ITP} \ll 1$ yields the linear proportionality of $h_{ITP} \approx pk_{off}t_{ITP}(C_{T0}/K_D)$.[37] At the LOD concentration, signal strength, i.e. fraction of the bound probe, is the same for LF and ITP-LF. Therefore, $LOD_{LF} \approx h_{LOD}K_D$ and $LOD_{ITP-LF} \approx h_{LOD}/pk_{on}t_{ITP}$. As a result, the LOD ratio scales as $pk_{off}t_{ITP}$ which confirms our model predictions presented in Figure 4 that the LOD ratio is directly proportional to the ITP parameter $pt_{ITP}$ and the off-rate constant while it is independent of the $K_D$ value.

Our scaling analysis and the analytical model show that the amount of captured target for conventional LF assay depends on the normalized initial target concentration, i.e. $C_{T0}/K_D$. On the other hand, for ITP-LF the amount of captured target depends on $pk_{off}t_{ITP}(C_{T0}/K_D)$ which includes both antibody-specific parameters, $k_{off}$ and $K_D$, and an ITP condition parameter $pt_{ITP}$. This offers flexibility in assay design by controlling the preconcentration ratio and the ITP residence time. For a point of care LF test, known values can be kinetic parameters of the antibody-antigen pair and the desired test time. From this, one can estimate the ITP preconcentration ratio and residence time required to achieve the desired detection limits. Preconcentration ratio in ITP can be optimized by carefully selecting the electrolyte system, optimizing the applied electric field and the device length.[40]

## 5. Summary

We demonstrate LOD and assay time improvement of LF assays using ITP. Our method leverages high preconcentration power of ITP to overcome the slow reaction kinetics of surface binding in LF. We focused target antibodies into a narrow ITP zone (order 100 μm) and transported them across the nitrocellulose membrane to the test zone. Our quantitative analysis show that LOD of LF assays can be improved by 400-fold, 160-fold and 60-fold at assay time of 90 sec, 5 min and 1 hour, respectively. Our qualitative colorimetric measurements show about 100-fold improvement in LOD using ITP, consistent with our quantitative findings. We used an analytical model describing the ITP-enhanced surface reaction in LF assays which confirms our experimental observations and serves as guide toward the design of LF assay enhanced by ITP. By integrating ITP into LF assay we can also capture up to 30 % which shows more than one order of magnitude improvement over conventional LF assays. High target extraction and preconcentration of ITP enables the LF assays to use large volume of sample when the target is very dilute, e.g. blood, and waste water. Moreover, we found that detection limit of LF can be improved by separating and preconcentrating target analytes from 5x diluted mouse serum (data provided in SI) showing the potential of ITP to improve LF with real-world complex biological samples.

**Associated Content**

Videos showing isotachophoresis-enhanced lateral flow assay (ITP-LF) using Alexa Fluor 488 (Video S-1) and colloidal gold (Video S-2) labelled IgG, electrophoretic mobility measurements of IgG secondary antibody, an example of calibration curves used for fluorescence quantitative measurements, target capturing in conventional LF assay, a discussion on surface reaction rate speed-up by ITP preconcentration and demonstration of ITP-LF in complex samples. are available as supporting information. This material is available free of charge via the Internet at http://pubs.acs.org.

**Author Information**

* E-mail: jposner@uw.edu, Phone: (206) 543-9834

**Notes:** The authors declare no competing financial interest.

**Acknowledgments**

This work was sponsored by an NSF CAREER Award (J.D.P., grant number CBET-0747917). We thank Carly Holstein and Paul Yager from Bioengineering department of the University of Washington for assistance.

**Figures and tables**

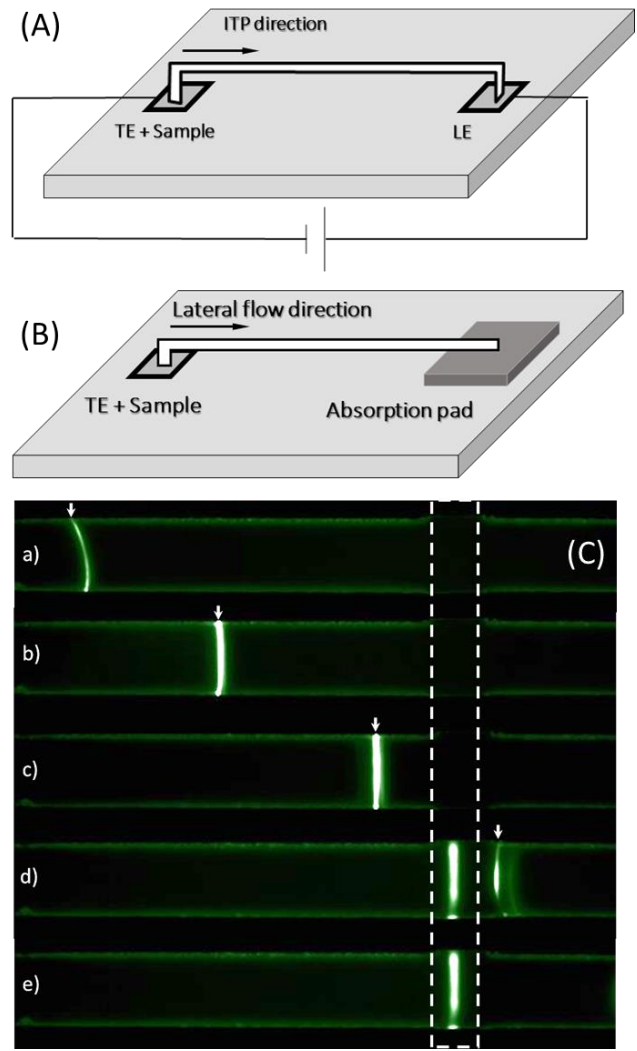

**Figure 1.** Experimental setup for (A) ITP-enhanced LF and (B) conventional LF assay. For both assays a 40 mm long nitrocellulose membrane is used as the paper device. In ITP-LF targets migrate electrophoretically under the high electric field of 500 µA and accumulate at the narrow interface of the TE and LE. In LF assay targets migrate through the device by the capillary action. High wicking properties of the absorption pad guarantees continuous flow of the sample. (C) Experimental snapshots taken at 5 different times showing ITP-focused IgG labeled with AF488 is captured at the test zone. (a-c) ITP plug forms and migrates on the membrane, (d)

current is reduced to 50 µA while the ITP zone crosses the test zone to increase the reaction time (e) after the plug is passed the test zone we see a specific fluorescence signal is generated. The white arrow indicates instant location of the ITP plug.

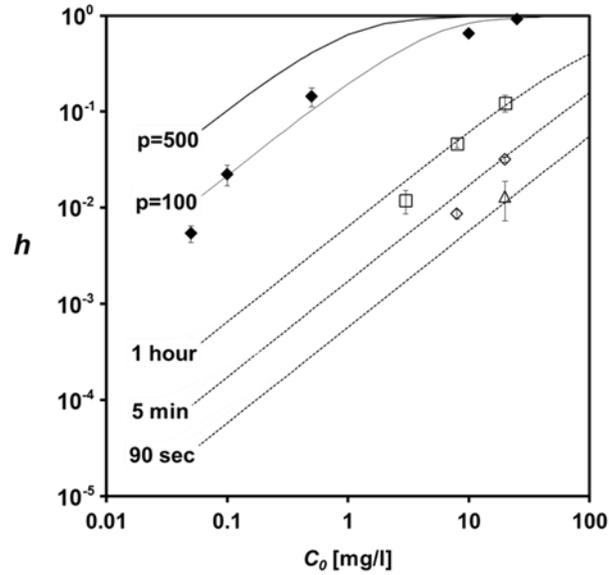

**Figure 2.** Experimental data showing quantitative detection of IgG using ITP-LF (closed diamonds) and its comparison to conventional LF at 90 second (open triangle), 5 minutes (open diamonds), and 1 hour (open squares). Along with experimental data (symbols), we show results of analytical models for the conventional (dashed) LF and for the ITP-LF (solid) assays. ITP creates two orders of magnitude improvement in LOD of LF after 7 minutes assay time. To generate the model curves we set the $k_{off} = 1.75 \times 10$-3 s$^{-1}$ and $K_D = 1.42 \times 10$-6 M , $p = 92$ and $t_{ITP} = 284$ s. The error bars indicate 95% confidence interval for three measurements.

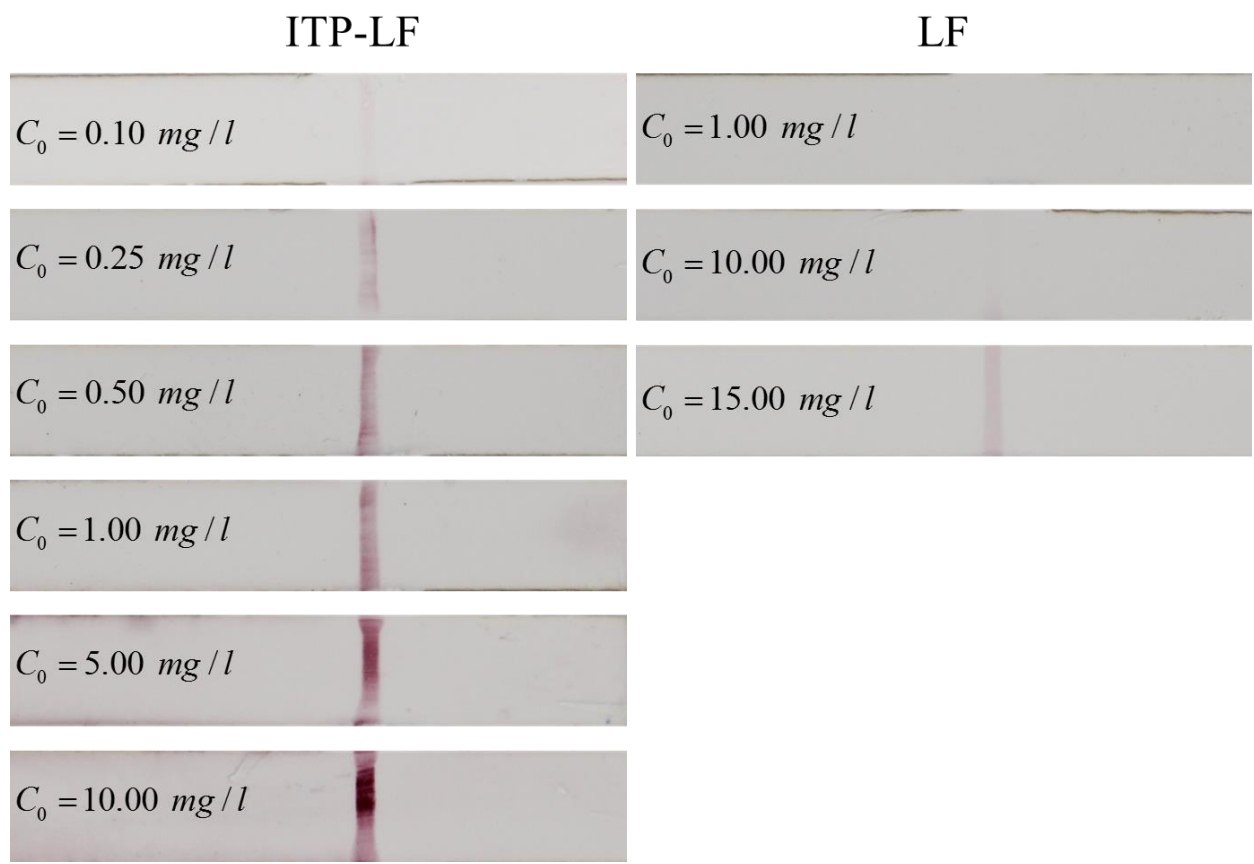

**Figure 3.** Qualitative detection limit of ITP-LF (right) compared to LF (left). Here the target is Goat anti Mouse IgG labeled with 40 nm colloidal gold. LF assays are performed for 5 min to be consistent with ITP assay time.

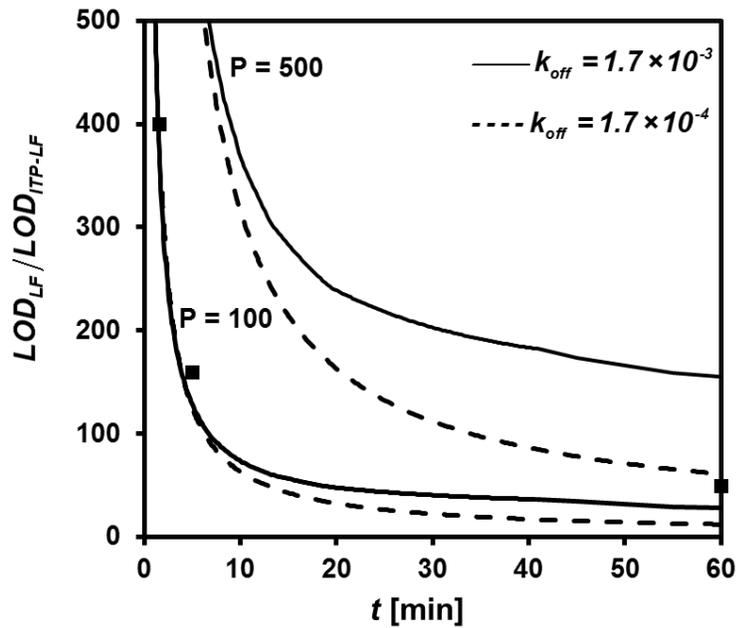

**Figure 4.** Ratio of $LOD_{LF}/LOD_{ITP-LF}$, predicted by our analytical model, versus the assay time, $t$. Each value on y-axis shows the level of improvement in LOD of LF. Contours are plotted for $t_{ITP}$ = 284 s, $p$ = 100 (bottom), $p$ = 500 (top), $k_{off}$ = 1.75×10$^{-3}$ s$^{-1}$ (solid lines) and $k_{off}$ = 1.75×10$^{-4}$ s$^{-1}$ (dashed lines). We note that contours have no dependence on the $K_D$ value showing that ITP improves detection limit of LF by promoting the reaction rate. Solid symbols represent our experimental conditions where $p$ = 92, $t_{ITP}$ = 284 s and different LF assay time showing good agreement with the model prediction.

**Table 1.** Detection limit improvement of conventional LF assay using ITP.

| Assay time | $LOD_{LF}$ (mg/l) | $LOD_{ITP\text{-}LF}$ (mg/l) | Fold improvement of detection limit |
|---|---|---|---|
| 90 seconds | 20 | 0.05 | 400 |
| 5 minutes | 8 | 0.05 | 160 |
| 1 hours | 3 | 0.05 | 60 |

**Table 2.** Fraction of captured target, $N_C/N_{T0}$, by ITP-LF compared to that of LF. $C_{T0}$ is the initial target concentration, $N_{T0}$ is the initial number of moles of target, and $N_{P0}$ is the number of moles of probe on the surface. The uncertainties represent 95% confidence interval for three measurements.

| $C_{T0}$ [a] | $N_{P0}$ [b] | $(N_{T0})_{ITP-LF}$ [b] | $(N_C/N_{T0})_{ITP-LF}$ [c] | $(N_{T0})_{LF}$ [b] | $(N_C/N_{T0})_{LF}$ [c] |
|---|---|---|---|---|---|
| 0.05 | 0.5 | 0.02 | 11.62 ± 2.29 | 0.06 | No detection |
| 0.1 | 0.5 | 0.04 | 24.05 ± 5.83 | 0.13 | No detection |
| 0.5 | 0.5 | 0.23 | 30.86 ± 6.91 | 0.67 | No detection |
| 10 | 0.5 | 4.6 | 7.01 ± 0.32 | 10.6 | 0.47 ± 0.05 |
| 25 | 0.5 | 11.6 | 3.94 ± 0.08 | 26.6 | 0.70 ± 0.04 |

[a] in mg/l, [b] in fmoles, [c] in %

**For TOC only**

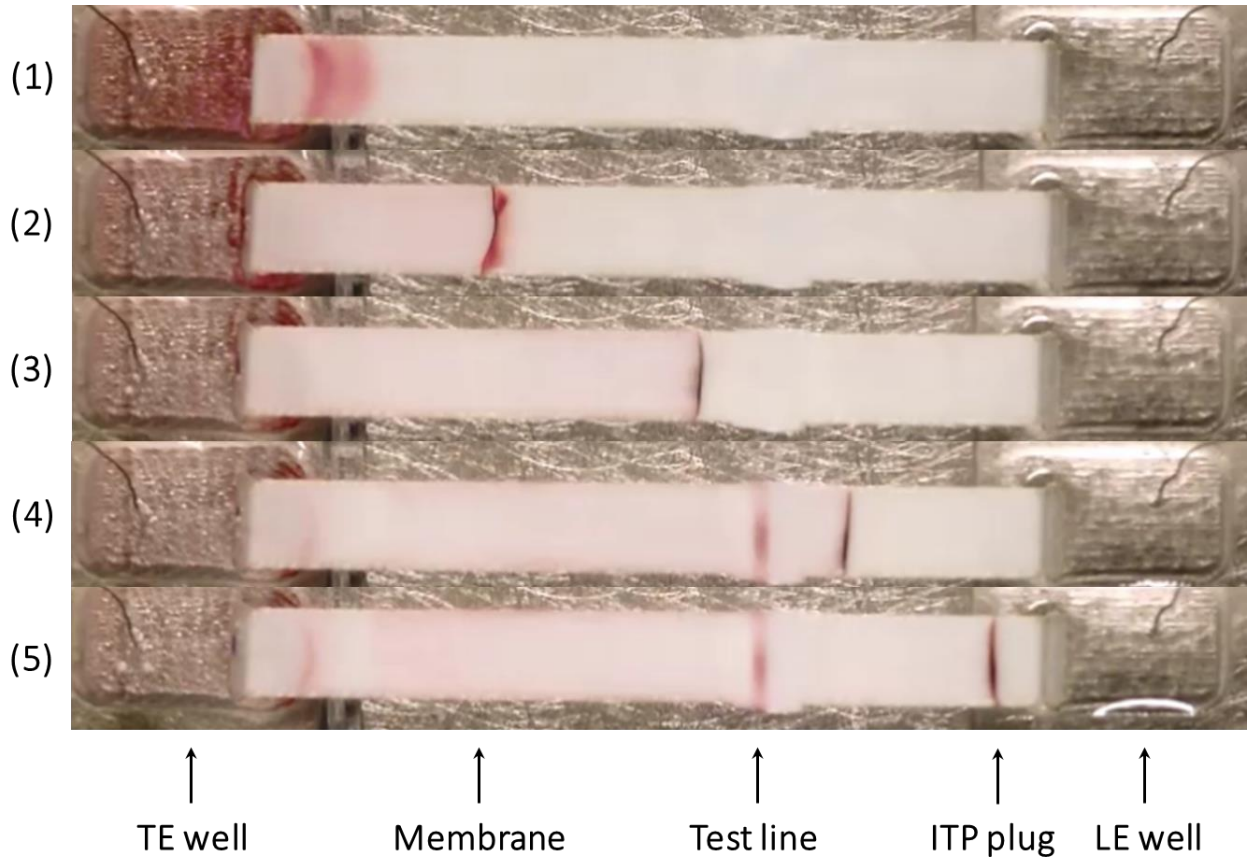

TE well    Membrane    Test line    ITP plug    LE well

Supporting Information

# Two orders of magnitude improvement in detection limit of lateral flow assays using isotachophoresis

*Babak Y. Moghadam†, Kelly T. Connelly‡, Jonathan D. Posner†\**

†University of Washington, Mechanical Engineering Department, Seattle, WA 98195, USA

‡University of California, Los Angeles, Mechanical and Aerospace Engineering Department, Los Angeles, CA 90095, USA

\*Corresponding author: Phone: (206) 543-9834, E-mail: jposner@uw.edu

**Abstract**

Supporting information provides two videos showing isotachophoresis-enhanced lateral flow assay (ITP-LF) using Alexa Fluor 488 (Video S-1) and colloidal gold (Video S-2) labelled IgG as the target, electrophoretic mobility measurements of IgG secondary antibody, an example of calibration curves used for fluorescence quantitative measurements, target capturing in conventional LF assay, a discussion on surface reaction rate speed-up by ITP preconcentration and demonstration of ITP-LF in complex samples.

**S-1. Electrophoretic mobility of IgG**

To design the ITP electrolyte system for ITP-enhanced LF we first measured the electrophoretic mobility of our target, goat anti-rabbit IgG labelled with Alexa Fluor 488, in a wide range of pH using Malvern Zetasizer Nano-Zs, Westborough, MA. We performed the measurements at 10 mM ionic strength which is close to the ionic strength of the trailing electrolyte (TE). Figure S-1 shows the measured mobility of IgG as a function of pH and its comparison to the mobility values of HEPES and Glycine calculated using the theoretical models.[1] IgG antibodies are large molecules of about 150 kDa and they show low electrophoretic mobilities especially around the physiological pH. Based on the measured IgG mobility of 6.3 nm$^2$/Vs at pH = 7.4 we chose Glycine as the TE ions and designed TE composition as 10 mM Glycine, with the mobility of 0.4 nm$^2$/Vs, buffered with Bis-Tris to pH 7.4. At this pH the binding reaction occurs at the neutral condition and the difference between mobilities of the target and TE ions is maximized which is preferred for more effective sample focusing and minimal sample dispersion.

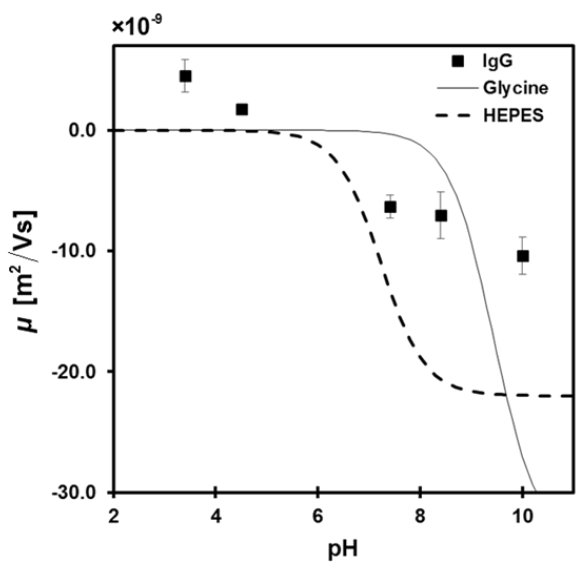

**Figure S-1.** Experimentally determined mobility of IgG as a function of pH. These values are compared with the theoretical predictions for HEPES and Glycine showing that at pH values higher than isoelectric point of IgG, Glycine can be used as the trailing ions while using, for example, HEPES does not result in an isotachophoretic preconcentration.

**S-2. Calibration curves for IgG labeled with Alexa Fluor 488**

We used Alexa Fluor 488 labeled IgG as the target to perform quantitative LF and ITP-LF assays. The fluorescence intensity of the target should be converted to concentration in order to calculate the amount of the target focused in the ITP plug and captured at the test zone. We performed calibration experiments before running each set of LF and ITP-LF experiments each day. For each calibration experiment we fully wet the membrane with a known concentration of the labeled target and measured fluorescence intensity of the membrane at three different molar concentrations of 10 µM, 25 µM, and 50 µM. For our quantitative analysis, we calculate the sample concentration, $C_{target}$, as:

$$C_{target} = \frac{C_{dye}}{I_{dye} - I_{back}} \left( I_{target} - I_{back} \right) \tag{S-1}$$

where $I_{target}$ is the fluorescence intensity of the captured/focused target, $I_{dye}$ is the signal intensity corresponding to a known concentration of highly concentrated labeled target, $C_{dye}$, and $I_{back}$ is the background intensity measured when the membrane is fully wetted by the buffer, *i.e.* zero target concentration. In practice, $C_{dye}/(I_{dye} - I_{back})$ is the slope of the linear regression fit to the calibration points using the least square method. Figure S-2 shows an example of our calibration curves, produced at five different days for three concentrations, demonstrating high consistency and reproducibility of our calibration curves. We used slope of the regression line fitted to these experimental data to convert the intensity values to the target concentration.

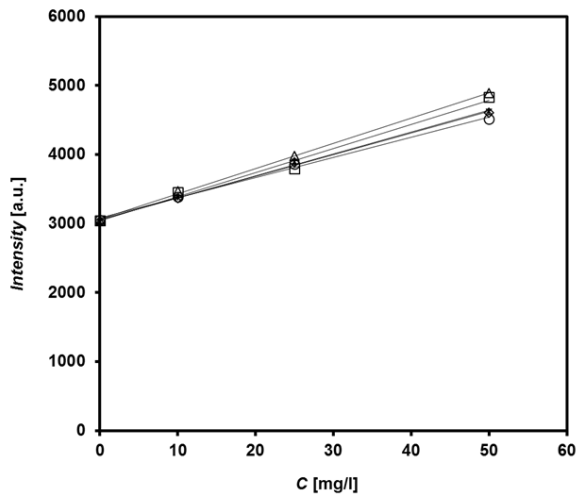

**Figure S-2.** Fluorescence intensity of the paper device wetted by a known concentration of the labeled target. These calibration points are used to convert fluorescence intensity of the target to concentration. Symbols represent the experimental data and solid lines are the fitted linear regression. Data are shown for the measurements performed at 5 different days showing high consistency and reproducibility of the calibration curves.

**S-3. Target capturing in conventional LF assay**

Figure S-3 shows the captured amount of target in LF assay as a function of time. We performed a conventional LF assay experiment using initial target concentration, $C_{T0}$, of 25 mg/l and monitored the amount of target captured, $m_C$, in the test zone over time. As predicted by Equation 1, the amount of captured target increases with time and reaches the value of 200 ng after 30 minutes. The rate at which the target is captured in LF assay depends mainly on the off-rate constant of the reaction, $k_{off}$. Moreover, the absolute amount of target captured at each time depends on the normalized initial target concentration, $C_{T0}/K_D$. Therefore we fit Equation S-1 to the experimental data provided in Figure S-3 to calculate the $k_{off}$ and $K_D$ as the fitting parameters.

$$\frac{m_C}{m_{P0}} = \frac{C_0^*}{C_0^* + 1}\left(1 - \exp\left(-\left(C_0^* + 1\right)k_{off} t\right)\right) \tag{S-1}$$

Here $m_C$ and $m_{P0}$ are respectively mass of the captured target and the mass of the probe at $t = 0$. $m_C$ is obtained directly by converting fluorescence intensity of the target in the test zone and to the mass of the captured target. $m_{P0}$ is determined by knowing the flow rate at which the captured reagent is spotted on the membrane and its initial concentration. $C_0^*$ is defined as normalized initial target concentration, $C_{T0}/K_D$ where $C_{T0}$ is the molar concentration of the suspended target in the immediate vicinity on the immobilized probe in the test zone. We derived Equation S-1 by replacing $C_C$ and $C_{P0}$ in Equation 1 with $m_C$ and $m_{P0}$, respectively, assuming same molecular weight for the target and probe IgGs and their one-to-one binding. In Figure S-3 we calculated the fitting parameters as, $k_{off} = 1.75 \times 10^{-3}$ s$^{-1}$ and $K_D = 1.42 \times 10^{-6}$ M which are consistent with the range of values reported for the interaction of secondary antibodies.[2,3]

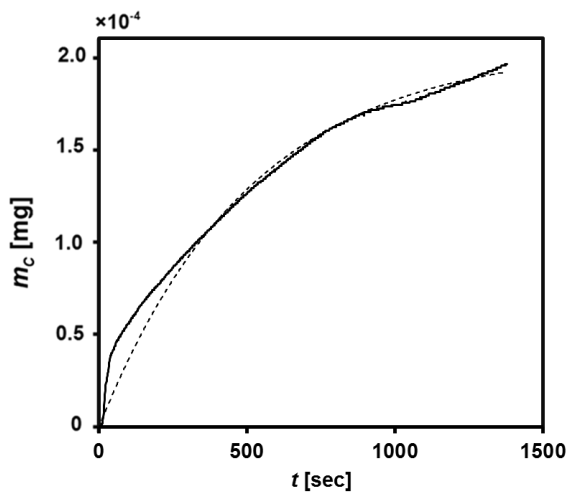

**Figure S-3.** Amount of sample captured in a conventional LF assay as a function of time. Dotted line represents the theoretical model curve fitted to the experimental data to determine kinetic parameters of the target-probe pair as $k_{off} = 1.75 \times 10^{-3}$ s$^{-1}$ and $K_D = 1.42 \times 10^{-6}$ M.

## S-4. Increasing reaction rate and signal amplification

Analytical model developed by Han et al., described in Equations 1 and 2, predicts that ITP preconcentration can enhance LF by increasing the reaction rate.[4] Increasing the binding reaction rate by ITP preconcentration scales with $p$ (e.g., which can be seen by taking the derivative of Equation 2 with respect to $t_{ITP}$) compared to kinetically limited LF without preconcentration. We compare the time to obtain the same signal intensity from both methods. We calculate 99% characteristic time, i.e., the time required to reach 99% of the equilibrium signal, $h_{eq}$, defined as $C_0^*/(C_0^*+1)$ for a given target concentration. In Figure S-4, dashed contour lines represent the ratio of kinetically limited LF assay time to ITP-LF assay time. For a common range of $p$ and $C_0^*$ values presented here, we see that an order of 500-fold speedup is predicted. Moreover, we predict an increase in detection limits, i.e. signal amplification, for ITP-LF.

In Figure S-4 solid contour lines represent the ratio of the fraction of bound probe with and without ITP versus the nondimensional time, $k_{off}t_{ITP}$, and the preconcentration level, $p$. For the solid lines, we set $C_0^*$ at $6.67 \times 10^{-2}$ (corresponds to $C_0 = 10$ mg/l). For example, for a typical $k_{off}$ value of order $10^{-3}$ s$^{-1}$, residence times of 10 sec or greater result in $h_{ITP}/h_{eq}$ values greater than unity for preconcentration factors of about 100 or greater. Such preconcentration values and residence times are easily achievable with ITP for a wide variety of targets and sample types. We note that the y- and x-axis of Figure S-4 are not completely orthogonal since $p$ and $t_{ITP}$ are both functions of the electric field. Han et al. showed this representation because these dimensionless parameters facilitate comparison with experimental conditions. The square symbol in Figure S-4 represent our experimental conditions *$p = 92$, $t_{ITP} = 284$ s, and $k_{off} = 1.75 \times 10^{-3}$ s$^{-1}$*. The model predicts about 13-fold increase in the fraction of bound probe using ITP at $C_0 = 10$ mg/l consistent with our experimental measurement presented in Figure 2 which shows 14.1 fold increase in the fraction of bound probe. Figure S-4 also predicts about 400-fold decrease in time for our LF reaction using ITP. However, our experimental observations suggest that the reaction speed up factor is 12, considering ITP residence time of 284 and LF assay time of 1 hour (LF equilibrium time). This discrepancy is due the fact that we reduce the current to 50 µA when the ITP zone reaches the test zone. For 500 µA applied current, we calculated $t_{ITP}$ = 9 sec and results in the reaction speed up of 400 which confirms our model predictions. Figure S-4 shows that in addition to improving LOD of LF, ITP can shorten the assay time which is an important criteria for POC testing.

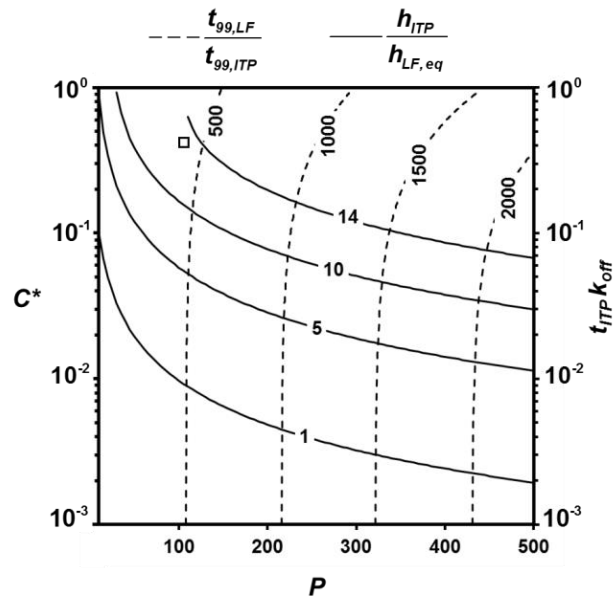

**Figure S-4.** Comparison of signal amplification and reaction time between ITP-LF and kinetically limited LF. Dashed contours represent the ratio of characteristic time to reach 99% of the equilibrium (after 1 hour) with LF to that with *ITP, $t_{99,LF}/t_{99, ITP}$*, for a common range of $C_0^*$ and $p$. We defined $C_0^*$ as a dimensionless parameter, $C_0/K_D$. Solid contour lines represent the ratio of the fraction of bound probes in ITP-LF to that from the LF at equilibrium, $h_{ITP}/h_{LF,eq}$, and are plotted for relevant values of nondimensional time $t_{ITP}k_{off}$ and $p$. For the solid lines, we set $C_0^*$ at $6.67 \times 10^{-2}$ (corresponds to $C_0$ = 10 mg/l). The square symbol represents the model

prediction for the estimated experimental conditions used in Figure 2.

**S-5. Analyte extraction from a complex sample using ITP**

In order to show capability of ITP to extract target from a complex sample on paper-based devices and improve LOD of LF assays, we performed ITP-LF by dispensing our target IgG labeled with Alexa Fluor 488 in mouse blood serum. Blood serum includes all proteins not used in blood clotting (coagulation) and all the electrolytes, antibodies, antigens, hormones, and any exogenous substances (e.g., drugs and microorganisms), therefore it is a biological complex sample. We followed the same experimental conditions and ITP electrolyte system as described for regular ITP-LF. The sample reservoir was contained the TE mixed with 5 mg/l of the target dispersed in 5x and 10x blood serum. We were able to successfully separate and focus our target sample from the blood serum and transport it to the test zone while no detection observed using LF at these concentrations, showing detection limit improvement of LF assay using ITP in complex samples. Our quantitative measurements presented in Table S-1 show that the fraction of bound probe using 5x and 10x serum is $h = 18$ % which is close to binding ratio of $h = 34$ % obtained from regular ITP-LF, at the same target initial concentration. We observed the same trend for the extraction ratio which is 3.8 % for the ITP-LF in complex matrix and is 7.3% for regular ITP-LF. The lower binding values observed using the complex sample may be due to non-specific binding of our target IgG with the IgG molecules present in the blood serum and the altered properties of the TE solution. Adding blood serum to the TE may change pH of the solution which has a great effect on the effective mobility of the TE ions. We hypothesize by optimizing chemical composition of the TE based on the complex matrix we are able to reach binding ratios close to the regular ITP-LF experiments.

**Table S-1.** Binding and capturing ratios obtained from ITP-LF performed in mouse serum as the complex sample and its comparison with regular ITP-LF.

| Sample | Initial conc. (mg/l) | Binding ratio (%) | Capture ratio (%) |
|---|---|---|---|
| IgG in DI | 5 | 0.34 ± 0.03 | 7.3 ± 0.3 |
| IgG in 5x serum | 5 | 0.18 ± 0.01 | 3.8 ± 0.3 |
| IgG in 10x serum | 5 | 0.18 ± 0.04 | 3.8 ± 0.8 |